\documentstyle[12pt]{article}
\renewcommand
\baselinestretch{1.4}

\begin{document}

\hfill{UM-P-93/54}

\hfill{OZ-93/12}
\vskip .5cm
\begin{center}

{\LARGE \bf
A Partial Unification Model in Non-commutative Geometry}

\vspace{4mm}

\renewcommand
\baselinestretch{0.8}\vspace{4mm}
{\sc B. E. Hanlon and G. C. Joshi}
 \\
{\it
Research Centre for High Energy Physics \\
School of Physics \\
University of Melbourne \\
Parkville, Victoria 3052 \\
Australia}

\renewcommand
\baselinestretch{1.4}

\vspace{5mm}

\end{center}
\begin{abstract}
We consider the construction of $SU(2)_{L}\otimes SU(2)_{R}\otimes
SU(4)$ partial unification models as an example of phenomenologically
acceptable unification models in the absence of supersymmetry in
non-commutative geometry. We exploit the Chamseddine, Felder and
Fr\"ohlich generalization of the Connes and Lott model building
prescription. By introducing a bi-module structure and appropriate
permutation symmetries we construct a model with triplet Higgs fields
in the $SU(2)$ sectors and spontaneous breaking of $SU(4)$.
\end{abstract}

\vspace{1cm}



\newpage

\section{Introduction}

As with other extensions of space-time, non-commutative geometry
provides a framework in which scalar Higgs fields may be introduced on
the same level as gauge fields. In higher dimensional models, Higgs
fields result from gauge fields which originally carried space
indicies corresponding to the, now compactified, additional dimensions.
While this procedure has an asthetic appeal, phenomenological problems
arise from the existence of a single order parameter associated with
the compactification scale, usually taken at the Planck scale{\cite A}.
Non-commutative geometry provides an alternative framework in which
differing scales may exist.

These geometrical considerations emerge from applications of gauge
theory beyond Riemannian spaces. The notion of a manifold is
generalized to be the product of a continuous manifold by a discrete
set of points. Gauge fields now arise from appropriately chosen fibre
bundles along the continuous directions, while Higgs scalars result
from gauging the discrete directions. Since spinor fields are the
fundamental fields in non-commutative gauge theory, the fermionic
action can be introduced in a simple way. Consequently, realistic
phenomenological models can be considered within this model building
prescription and indeed the standard model has been made the subject
of this approach{\cite B}.

If these notions are to be applied to GUT models then the original
model building prescription of Connes and Lott must be modified{\cite
{B,C}}. This
allows for gauge theories which are not constrained to be product
symmetries. The modification, introduced by Chamseddine, Felder and
Fr\"ohlich{\cite D},
consists of embedding the symmetry breaking in the Dirac
operator such that gauge invariance is not broken. This simplifies and
generalizes model building and allows for the introduction of
permutation symmetries between copies of space-time, yielding Higgs
representations necessary for symmetry breaking at appropriate scales.

The $SU(5)$ GUT model constructed by this approach did not provide any
additional suppression on the rate of proton decay and therefore is
ruled out experimentally. Appeals to space-time supersymmetry in
non-commutative geometry have yet to be formally developed and do not
appear to have an obvious answer. For this reason other avenues must
be explored in order to yield acceptable models. Such examples are
provided by GUT models with extended symmetry breaking schemes, such
as $SO(10)$. As well as suppression of the proton decay rate, such
quark-lepton unified models allow for the consistent inclusion of a
right handed neutrino and the freedom to incorporate other
phenomenological features, such as a reasonable value for
${\sin^{2}}\theta_{W}${\cite E}.
Originally it was speculated that the Higgs
fields required to implement such a scheme within non-commutative
geometry could not be easily constructed within the model building
prescription and would need to be added as an external field, not
associated to any vector{\cite D}.
Recently, however, Chamseddine and Fr\"ohlich
succeeded in constructing a consistent $SO(10)$ model{\cite F}.
This represents
an important step in the development of a deeper understanding behind
the possible origin of mass scales in such extended models by
generalizing the permutation symmetry between space-times to include
conjugation symmetries as well as direct identifications. In order to
realize an acceptable model, however, it was necessary to introduce
additional singlet spinors so that Higgs fields transforming as ${\bf
16}$'s could be included, yielding Cabibbo angle mixing among down
quarks.

While the $SO(10)$ breaking scheme is not unique, the spontaneous
breaking pattern realizing the Pati-Salam partial unification
$SU(2)_{L}\otimes SU(2)_{R}\otimes SU(4)$ has an appealing symmetry in
which the phenomenological featues are most transparent{\cite G}.
Importantly,
this is also the minimal symmetry group incorporating both
quark-lepton unification and the quantization of electric charge.
The
electroweak sector of such left-right symmetric theories has already
been investigated by Chamseddine et al. yielding spontaneous symmetry
breaking of $SU(2)_{R}$ by triplets{\cite D}. Other approaches to
non-commutative geometry, originally considered by Coquereux et
al.{\cite H},
Dubois-Violette et al.{\cite I} and Balakrishna et al.{\cite J},
have also been
generalized to yield a left-right symmetric weak interaction model,
this time with doublet Higgs fields{\cite K}.
It was found that this made
maximal use of the gauge connection in the discrete directions.
However, as well as lacking the same compelling geometrical structure,
it is difficult to introduce fermions in a straight forward way in
these alternative approaches, so restricting considerations to the
bosonic sector. Nevertheless, left-right symmetric models appear to
have a natural identification within non-commutative geometry.

In this paper we wish to extend this investigation of left-right
symmetric models to the Pati-Salam unification symmetry. We will
consider firstly a minimal model and then explicitly construct a model
with triplet Higgs in the $SU(2)$ sectors and spontaneous breaking of
$SU(4)$. To realize such a scheme we will include a bi-module
structure similar to that for the addition of $SU(3)$ colour to the
standard model{\cite B}.
However, unlike in the $SU(3)$ case, we do not wish
the symmetry breaking matrices in these directions to be identically
zero, thus introducing a non-trivial extension. We find that a very
natural model emerges without the need to introduce singlet spinors or
an additional set of conjugate fermions to produce coupling to
conjugate bi-doublet Higgs.

\section{The Model Building Prescription}

We wish to give an overview of the model building prescription
outlined by Chamseddine et al. {\cite{D}}. The geometrical setting
is that of Connes{\cite C}, with the modification
that the choice of gauge
structure is defined directly within the Dirac operator. This is in
contrast to the original prescription of Connes and Lott{\cite B}
where the
gauge structure results from the choice of vector bundle $\cal E$,
defined as a finite projective right module over the algebra $\cal A$
defining the non-commutative space. It follows that the natural choice
of vector bundle must now be ${\cal E} = {\cal A}$ i.e. the orthogonal
projection is trivial. With this choice the connection and curvature
have the simplest form.

The notion of geodesic distance is incorporated in the concept of a
$K$-cycle. A $K$-cycle over the involutive algebra $\cal A$ is a
$\ast$-action of $\cal A$ by bounded operators on a Hilbert space
$\cal H$
and a possibly unbounded, self-adjoint operator $D$, denoted Dirac
operator, such that $[D,f]$ is a bounded operator for all $f\in {\cal
A}$ and $(1+ D^{2})^{-1}$ is compact. As we will be interested in four
point spaces we will cast definitions about this choice although they
are easily extended to any number of points. Let $X$ be a compact
Riemannian spin manifold, ${\cal A}_{1}$
the algebra of functions on $X$
and $({{\cal H}_{1}}, D_{1})$
the Dirac $K$-cycle with ${\cal H}_{1}= L^{2}(X,{\sqrt
g} d^{d} x)$ on ${\cal A}_{1}$. Denote by $\gamma_{5}$ the fifth
anticommuting Dirac gamma matrix, the chirality operator, given by
$\gamma_{5} =\gamma_{1} \gamma_{2} \gamma_{3} \gamma_{4}$ in four
dimensional Euclidean space, defining a $Z_{2}$ grading on ${\cal
H}_{1}$. Let ${\cal A}_{2}$ be given by ${{\cal A}_{2}}
= M_{n}(C)\oplus
M_{p}(C)\oplus M_{q}(C)\oplus M_{r}(C)$, where $M_{n}(C)$ is the
set of all $n\times n$ matrices, with the $K$-cycle $({{\cal H}_{2}},
D_{2})$ and ${{\cal H}_{2}} = {{\cal H}_{n}}\oplus {{\cal H}_{p}}\oplus
{{\cal H}_{q}}\oplus {{\cal H}_{r}}$
corresponding to the Hilbert spaces
$C^{n}, C^{p}, C^{q}$ and $C^{r}$ respectively. The product geometry
is then given by
\begin{equation}
{\cal A} = {\cal A}_{1}\otimes {\cal A}_{2} \;\; ,
\end{equation}
with the Dirac operator correspondingly written as
\begin{equation}
D = D_{1}\otimes 1\otimes 1 + \gamma_{5} \otimes D_{2} \;\; .
\end{equation}
The decomposition of ${\cal H}_{2}$ diagonalizes the action of $f\in
{\cal A}$
\[ f \rightarrow {\rm diag}(f_{1}, f_{2}, f_{3}, f_{4}) \;\; .\]
The operator $D$ is then
\begin{equation}
D=
\left (
\begin{array}{cccc}
\;\;\;\; {\not\! \partial}\otimes 1\otimes 1 \;\;\;\;
\;\;\; \gamma_{5}\otimes
M_{12}\otimes K_{12} \;\;\; \gamma_{5}\otimes M_{13}\otimes K_{13}
\;\;\; \gamma_{5}\otimes M_{14}\otimes K_{14} \;\; \\
\;\gamma_{5}\otimes M_{21}\otimes K_{21} \;\;\;
\;\;\;{\not\!\partial}\otimes 1\otimes 1 \;\;\;\;\;\;\;\;
\gamma_{5}\otimes M_{23}\otimes K_{23} \;\;\;
\gamma_{5}\otimes M_{24}\otimes K_{24} \;\; \\
\;\gamma_{5}\otimes M_{31}\otimes K_{31} \;\;\;
\gamma_{5}\otimes M_{32}\otimes K_{32} \;\;\;
\;\;\;{\not\! \partial}\otimes 1\otimes 1 \;\;\;\;\;\;\;\;\;
\gamma_{5}\otimes M_{34}\otimes K_{34} \;\; \\
\;\gamma_{5}\otimes M_{41}\otimes K_{41} \;\;\;
\gamma_{5}\otimes M_{42}\otimes K_{42} \;\;\;
\gamma_{5}\otimes M_{43}\otimes K_{43} \;\;\;
\;\;\;{\not\! \partial}\otimes 1\otimes 1 \;\;\;\;\;\;\;\;
\end{array}
\right )
\end{equation}
where $M_{mn}$, $m\not= n$, is an $m\times n$ complex matrix such that
$M^{\dagger}_{mn} = M_{nm}$ and $K_{mn}$ are $3\times 3$ family mixing
matrices. The $M_{mn}$ correspond to the tree level vacuum expectation
values of Higgs fields, the chosen form of which determines the
symmetry breaking pattern.

The space of forms $\Omega^{\ast} ({\cal A}) = \oplus^{\infty}_{n=0}
\Omega^{n} ({\cal A})$ is generated by elements
$a_{0}da_{1}.....da_{k} \in \Omega^{k}({\cal A})$ such that $a_{0},
a_{1},..... \in {\cal A}$. With ${\cal E} = {\cal A}$ a
connection is given by the element
\begin{equation}
\rho = \sum_{i} a^{i} db^{i} \in \Omega^{1}({\cal A}) \;\; ,
\end{equation}
with the curvature specified by
\begin{equation}
\Theta = d\rho + \rho^{2} \in \Omega^{2}({\cal A})
\end{equation}
where $d1=0$ and the $\rho^{2}$ term does not vanish. An involutive
representation of $\Omega^{\ast} ({\cal A})$ by bounded operators on
$\cal H$, with algebra $B({\cal H})$, is defined by the map $\pi :
\Omega^{\ast} ({\cal A}) \rightarrow B({\cal H})$ given by
\begin{equation}
\pi (a_{0}da_{1}.....da_{n}) = a_{0}[D, a_{1}][D, a_{2}]....[D, a_{n}]
\;\; .
\end{equation}
Consequently
\begin{equation}
\pi (\rho ) = \sum_{i} a^{i}[D, b^{i}] \;\; .
\label{connrep}
\end{equation}
Evaluating (\ref{connrep}) yields the result
\begin{equation}
\pi (\rho ) =
\left (
\begin{array}{cccc}
\;\;\;\;\;\;\;\;
 A_{1} \;\;\;\;\;\;\;\;\;\;\;\;\,\,\,
 \gamma_{5}\otimes \phi_{12}\otimes K_{12}
\;\;\;
\gamma_{5}\otimes \phi_{13}\otimes K_{13} \;\;\;
\gamma_{5}\otimes \phi_{14}\otimes K_{14} \;\; \\
\;\gamma_{5}\otimes \phi_{21}\otimes K_{21} \;\;\;\;\;\;\;\;\;
 \;\;\; A_{2} \;\;\;\;\;\;\;\;\;\;\;\;\;
\gamma_{5}\otimes \phi_{23}\otimes K_{23} \;\;\;
\gamma_{5}\otimes \phi_{24}\otimes K_{24} \;\; \\
\;\gamma_{5}\otimes \phi_{31}\otimes K_{31} \;\;\;
\gamma_{5}\otimes \phi_{32}\otimes K_{32} \;\;\;\;\;\;\;\;\;
\;\;\;A_{3} \;\;\;\;\;\;\;\;\;\;\;\;\;
\gamma_{5}\otimes \phi_{34}\otimes K_{34} \;\; \\
\,\,\,\gamma_{5}\otimes \phi_{41}\otimes K_{41} \;\;\;
\gamma_{5}\otimes \phi_{42}\otimes K_{42} \;\;\;\,
\gamma_{5}\otimes \phi_{43}\otimes K_{43} \;\;\;\;\;\;\;
\;\;\;\; A_{4} \;\;\;\;
\;\;\;\;\;\;\;\;\;
\end{array}
\right )
\end{equation}
where the $A$'s and $\phi$'s are determined in terms of the $a$'s and
$b$'s by
\begin{eqnarray}
A_{m} & = & \sum_{i} a^{i}_{m} {\not \! \partial} b^{i}_{m} \;\; ,
\nonumber \\
\phi_{mn} & = & \sum_{i} a^{i}_{m}(M_{mn}b^{i}_{n} - b^{i}_{m}
M_{mn})
\end{eqnarray}
satisfying $A^{\dagger}_{m} = A_{m}$ and $\phi^{\dagger}_{mn} =
\phi_{nm}$. The two form $d\rho = \sum_{i} da^{i}db^{i}$ with image
under $\pi$ given by $\pi (d\rho )=\sum_{i} [D, a^{i}][D, b^{i}]$ can
be similarly evaluated.

Unitary gauge transformations by $g\in U({\cal A}) = \{ g\in {\cal A}
: g^{\dagger}g=1\}$ can be defined in terms of transformations on the
$a^{i}$ and $b^{i}$ such that
\begin{eqnarray}
a^{i} & \rightarrow & ^{g}a^{i} = ga^{i} \; , \nonumber \\
b^{i} & \rightarrow & ^{g}b^{i} = b^{i}g^{\dagger} \;\; .
\end{eqnarray}
This definition implies the constraint
\begin{equation}
\sum_{i} a^{i}b^{i} =1 \;\; ,
\end{equation}
which can be imposed without loss of generality. It is straightforward
to compute the action of gauge transformations on $\pi (\rho )$ which
in component form can be written as
\begin{eqnarray}
^{g}A_{m} & = & g_{m}A_{m}g^{\dagger}_{m} + g_{m}{\not \!
\partial}g^{\dagger}_{m} \; , \nonumber \\
^{g}(\phi_{mn} + M_{mn}) & = & g_{m} (\phi_{mn} +
M_{mn})g^{\dagger}_{n} \;\; .
\end{eqnarray}
Thus the $A_{m}$ are the gauge fields while $\phi_{mn} +M_{mn}$ are
scalar fields transforming covariantly. The $\phi_{mn}$ represent
fluctuations around the vacuum state so that we are in fact working in
the spontaneously broken phase for which the Higgs potential will be
minimized when $\phi_{mn}=0$.

A crucial aspect which must be considered is that the representation
$\pi$ is ambiguous, with the correct space of forms actually given by
${\Omega^{\ast}({\cal A})}/{Ker\pi + dKer\pi}${\cite B}. Working on $
\Omega^{\ast}({\cal A})$ will result in the appearance of auxiliary
fields into which the scalar Higgs potential could be absorbed,
removing the Higgs mechanism from the model. The potential is saved
from disappearing, however, by including the $3\times 3$ family mixing
matrices $K_{mn}$. Nevertheless, in calculating the potential it is
necessary to determine which of the auxiliary fields are truly
independent. If all the auxiliary fields are independent the Higgs
potential will disappear regardless. This places severe constraints on
model building and the choice of vacuum expectation values for which
the independence or otherwise of the auxiliary fields will depend.

Since the fermionic fields are the fundamental fields, the spinor
action can be expressed simply as
\begin{equation}
I_{\Psi} = <\Psi, (D+ \pi (\rho ))\Psi > \;\; .
\label{faction}
\end{equation}
To determine the Yang-Mills action the notion of Diximier trace must
be considered. The action is given by the positive definite expression
\begin{equation}
I = {1/8}Tr_{w}(\Theta^{2} |D|^{-4}) \;\; ,
\end{equation}
where the Diximier trace is defined by
\begin{equation}
Tr_{w}(|T|)={\rm lim}_{w}{1\over {\rm log} N}\sum_{i}^{N} \mu_{i} (T)
\end{equation}
for a compact operator $T$ and eigenvalues $\mu_{i}$ of $|T|$. For the
Dirac operator the action can be equivalently expressed as
\begin{equation}
I = {1/8} \int d^{4}xTr(tr(\pi^{2} (\Theta )))\;\; ,
\label{action}
\end{equation}
where $tr$ is over the Clifford algebra and $Tr$ is over the matrix
structure. Finally, the action is analytically continued to Minkowski
space.

\section{A Minimal Model}

By minimal, we are making reference to a model for which the simplest
Higgs sector can be constructed to implement the required symmetry
breaking scheme.
This is in
analogy to the minimal $O(10)$ model of Witten{\cite L}.
We will consider a
Riemannian spin manifold extended by four points with the algebra
given by
\begin{equation}
{\cal A}_{2} = M_{2}(C)\oplus M_{4}(C)\oplus M_{4}(C)\oplus
M_{2}(C) \;\; ,
\end{equation}
together with the permutation symmetry
\[ \;\;\;\;\;\;\;\;\;\;
a_{2}^{i} = a_{3}^{i} \;\;\;\;\;\;\;\;\;\; b_{2}^{i} = b_{3}^{i}
\;\; . \]
In this way the second and third copies are identified and we have
Higgs fields transforming in a self adjoint rather than a product
representation in this region. With this choice the vector potential
$\pi (\rho )$ becomes
\begin{equation}
\pi (\rho ) =
\left (
\begin{array}{cccc}
\; A_{L} \;\; \chi_{L} \;\; \chi_{L} \;\; \phi \; \\
\; \chi_{L}^{\dagger} \;\; A_{4} \;\; \Sigma \;\; \chi_{R}^{\dagger}\;
\\
\; \chi_{L}^{\dagger} \;\; \Sigma \;\; A_{4} \;\; \chi_{R}^{\dagger}
\; \\
\; \phi^{\dagger} \;\; \chi_{R} \;\; \chi_{R} \;\; A_{R} \;
\end{array}
\right )
\label{pot2}
\end{equation}
where the gauge fields $A=\gamma^{\mu} A_{\mu}$ are self adjoint
$n\times n$ gauge vectors, $\Sigma$ is a self adjoint $4\times 4$
scalar field (i.e. $\Sigma_{23} = \Sigma_{32} =
\Sigma_{32}^{\dagger}$), $\chi_{L}$ and $\chi_{R}$ are $2\times 4$
complex scalar fields and $\phi$ is a bidoublet scalar field. $A_{L},
A_{R}$ and $A_{4}$ are $U(2)_{L}, U(2)_{R}$ and $U(4)$ gauge fields
respectively.

Note that the Pati-Salam partial unification has no $U(1)$ symmetries.
Thus in the reduction from $U(2)_{L}\otimes U(2)_{R}\otimes U(4)$ to
$SU(2)_{L}\otimes SU(2)_{R}\otimes SU(4)$ we do not need to relate or
introduce $U(1)$ factors. To induce the reduction we impose the
constraint
\begin{equation}
Tr(A_{L} + A_{R}) = 2Tr(A_{4}) =0 \;\; ,
\end{equation}
reducing $U(2)_{L}\otimes U(2)_{R}$ to $SU(2)_{L}\otimes SU(2)_{R}$
and $U(4)$ to $SU(4)$. $\Sigma$ will now introduce spontaneous breaking
of $SU(4)$ to $SU(3)\otimes U(1)_{B-L}$, $\chi_{L,R}$ allows for
assymetric breaking of $SU(2)_{L,R}$ by appropriate choices of vacuum
expectation values while $\phi$ is responsible for symmetry
breaking at the electroweak scale.

Introducing the fermionic sector now poses a dilema. The multiplet
structure for one family is
\begin{equation}
\psi_{L,R} =
\left [
\begin{array}{cccc}
\; u_{r} \;\; u_{b} \;\; u_{g} \;\; u_{l}= \nu_{e} \; \\
\; d_{r} \;\; d_{b} \;\; d_{g} \;\; d_{l}= e^{-} \;
\end{array}
\right ]_{L,R}
\label{mulstr}
\end{equation}
where lepton number is identified as the fourth colour. The
representation structure with respect to
$SU(2)_{L}\otimes SU(2)_{R}\otimes SU(4)$ is $\psi_{L} = ({\bf 2, 1,
4})$ and $\psi_{R} = ({\bf 1, 2, 4})$. Clearly, there are no gauge
invariant couplings of these fields with the Higgs scalars $\chi_{L}$
and $\chi_{R}$. Coupling of fermions to the bidoublet field $\phi$ is
responsible for the generation of the usual quark and lepton masses
but the generation of a heavy right handed neutrino depends
on an extended interaction sector.
This neutrino can get such a mass only through mixing with exotic
fermions{\cite M}.
This is difficult to implement in this scheme beyond the
inclusion of fermionic singlets. Note also that we cannot appeal to
higher order effects as we are dealing with classical geometries.
A viable model without exotics must, therefore, induce the
required breaking and mass generation at tree level from Higgs scalars
corresponding to the standard fermions only. That is we
require a non-minimal model.

It is worthwhile pointing out that, with no identification between the
$SU(2)$ and $SU(4)$ sectors, the fermions corresponding to these
copies of space-time will represent different fields. That is, we
would have fermions transforming in the fundamental representation
$\bf 2$ of $SU(2)$ and independent fermions in the $\bf 4$ of $SU(4)$,
rather than with the multiplet structure (\ref{mulstr}). While it is
possible to write down an appropriate spinor $\Psi$, these
identifications would have to be imposed externally. This problem of
identification is directly related to the indifference of the $SU(4)$
fermion representations on their chirality.

\section{A Non-minimal Model}

An examination of the vector potential (\ref{pot2}) demonstrates that
it is not possible to have Higgs scalars transforming as a product
representation with one component the adjoint of a chosen symmetry.
This is a limitation imposed by the matrix structure. Clearly, an
extension is necesarry if such Higgs scalars are to exist. We will
again consider a Riemannian spin manifold extended by four points,
this time with the algebra
\begin{equation}
{\cal A}_{2} = M_{2}(C)\oplus M_{2}(C)\oplus M_{2}(C)\oplus
M_{2}(C) \;\; .
\end{equation}
The $U(4)$ sector is now introduced to the four point space by adding
the auxiliary algebra ${\cal B}_{2}$, with right action on $\cal
H$,
given by
\begin{equation}
{\cal B}_{2} = M_{4}(C)\oplus M_{4}(C)\oplus M_{4}(C)\oplus M_{4}(C)
\;\; .
\end{equation}
We make the same natural choice of vector bundle $\cal F = B$, where
${\cal B} = {\cal A}_{1} \otimes {\cal B}_{2}$. The physical Hilbert
space can now be written as
\begin{equation}
{\cal P} = {\cal E} \otimes {\cal H} \otimes {\cal F} \;\; ,
\end{equation}
that is we have introduced a bi-module. Writing ${\cal H}_{2}^{\cal
A}$ and ${\cal H}_{2}^{\cal B}$ for the Hilbert spaces corresponding
to the algebras ${\cal A}_{2}$ and ${\cal B}_{2}$ respectively, ${\cal
H}$ can be suggestively written as
\begin{equation}
{\cal H} = {\cal H}_{2}^{\cal A}\otimes L^{2}(S)\otimes {\cal H}_{2}
^{\cal B} \;\; .
\end{equation}
We will consider this as defining $U(2)\otimes U(4)$ gauge structure
on each of the four copies of space-time as there is no reason,
apriori, to assume that a single gauge symmetry only can be associated
with each copy.

Corresponding to this extension we write down the generalized
connection one-form
\begin{equation}
\rho = \sum_{i} a^{i}db^{i}\otimes 1 + 1\otimes \sum_{i} A^{i}dB^{i}
\;\; .
\end{equation}
Note that the first 1 is a $4\times 4$ unit matrix and the second a
$2\times 2$ unit matrix. We now introduce the Dirac operator
\begin{equation}
D = D_{1}\otimes 1\otimes 1\otimes 1 + \gamma_{5}\otimes D_{2} \;\; ,
\end{equation}
where $D_{2}$ is given by
\begin{eqnarray}
D_{2} =
\left (
\begin{array}{cccc}
 \;\;\;\;\;\;\;\;\;\;\;\; 0 \;\;\;\;\;\;\;\;\;\;\;\;\;\;
m_{12}\otimes M_{12}\otimes K_{12} \;\;
 m_{13}\otimes M_{13}\otimes K_{13} \;\;
 m_{14}\otimes M_{14}\otimes K_{14}  \\
 m_{21}\otimes M_{21}\otimes K_{21} \;\;
\;\;\;\;\;\;\;\;\;\;\; 0 \;\;\;\;\;\;\;\;\;\;\;\;\;\;\;
m_{23}\otimes M_{23}\otimes K_{23} \;\;\,
m_{24}\otimes M_{24}\otimes K_{24}  \\
 m_{31}\otimes M_{31}\otimes K_{31} \;\;
m_{32}\otimes M_{32}\otimes K_{32} \;\;
\;\;\;\;\;\;\;\;\;\; 0 \;\;\;\;\;\;\;\;\;\;\;\;\;\;\;\;
m_{34}\otimes M_{34}\otimes K_{34}  \\
 m_{41}\otimes M_{41}\otimes K_{41} \;\;\,
m_{42}\otimes M_{42}\otimes K_{42} \;\;
m_{43}\otimes M_{43}\otimes K_{43} \;\;
\;\;\;\;\;\;\;\;\;\;\; 0 \;\;\;\;\;\;\;\;\;\;\;\;\;
\end{array}
\right )
\end{eqnarray}
with $m_{mn}$ the tree level vacuum expectation values in the $U(2)$
sector, $M_{mn}$ the vacuum expectation values in the $U(4)$ sector
and $K_{mn}$ are $3\times 3$ family mixing matrices. We will consider
the construction of $\pi (\rho )$ for the general case first,
introducing the relevant permutation symmetries between space-times
once the form of the action has been established.

Since $d1=0$ we can re-express the connection $\rho$ as
\begin{equation}
\rho = \sum_{i} (a^{i}\otimes 1)d(b^{i}\otimes 1) + \sum_{i} (1\otimes
A^{i})d(1\otimes B^{i}) \;\; .
\end{equation}
The image of $\rho$ under $\pi$ is then
\begin{eqnarray}
\pi (\rho ) & = & \sum_{i} (a^{i}\otimes 1)[D, b^{i}\otimes 1] +
\sum_{i} (1\otimes A^{i})[D, 1\otimes B^{i}] \nonumber \\
 & = & \pi (\rho )_{1} + \pi (\rho )_{2} \;\; ,
\end{eqnarray}
where (suppressing the $\gamma_{5}$'s for brevity)
\begin{eqnarray}
\pi (\rho )_{1} =
\left (
\begin{array}{cccc}
 \;\;\;\;\;\;\;\;\;\;\; A_{2} \;\;\;\;\;\;\;\;\;\;\;
\phi_{12}\otimes M_{12}\otimes K_{12} \;\;
\phi_{13}\otimes M_{13}\otimes K_{13} \;\;
\phi_{14}\otimes M_{14}\otimes K_{14}  \\
 \phi_{21}\otimes M_{21}\otimes K_{21} \;\;
\;\;\;\;\;\;\;\;\, A_{2} \,\;\;\;\;\;\;\;\;\;\;\;\;\;
\phi_{23}\otimes M_{23}\otimes K_{23} \;\;
\phi_{24}\otimes M_{24}\otimes K_{24}  \\
\phi_{31}\otimes M_{31}\otimes K_{31} \;\;
\phi_{32}\otimes M_{32}\otimes K_{32} \;\;
\;\;\;\;\;\;\;\;\;\, A_{2} \,\;\;\;\;\;\;\;\;\;\;\;\;
\phi_{34}\otimes M_{34}\otimes K_{34}  \\
\phi_{41}\otimes M_{41}\otimes K_{41} \;\;
\phi_{42}\otimes M_{42}\otimes K_{42} \;\;
\phi_{43}\otimes M_{43}\otimes K_{43} \;\;
\;\;\;\;\;\;\;\;\; A_{2} \;\;\;\;\;\;\;\;\;\;\;\;
\end{array}
\right )
\end{eqnarray}
and
\begin{eqnarray}
\pi (\rho )_{2} =
\left (
\begin{array}{cccc}
\;\;\;\;\;\;\;\;\;\;\;  A_{4} \;\;\;\;\;\;\;\;\;\;\;
m_{12}\otimes\Phi_{12}\otimes K_{12} \;\;
m_{13}\otimes\Phi_{13}\otimes K_{13} \;\;
m_{14}\otimes\Phi_{14}\otimes K_{14}  \\
 m_{21}\otimes\Phi_{21}\otimes K_{21} \;\;
\;\;\;\;\;\;\;\;\,  A_{4} \,\;\;\;\;\;\;\;\;\;\;\;\;\;
m_{23}\otimes\Phi_{23}\otimes K_{23} \;\;
m_{24}\otimes\Phi_{24}\otimes K_{24}  \\
 m_{31}\otimes\Phi_{31}\otimes K_{31} \;\;
m_{32}\otimes\Phi_{32}\otimes K_{32}
\;\;\;\;\;\;\;\;\;\,\; A_{4} \;\,\;\;\;\;\;\;\;\;\;\;\;\;
m_{34}\otimes\Phi_{34}\otimes K_{34}  \\
 m_{41}\otimes\Phi_{41}\otimes K_{41} \;\;
m_{42}\otimes\Phi_{42}\otimes K_{42} \;\;
m_{43}\otimes\Phi_{43}\otimes K_{43} \;\;
\;\;\;\;\;\;\;\;\;  A_{4} \;\;\;\;\;\;\;\;\;\;\;\;
\end{array}
\right )
\end{eqnarray}
with $\phi_{mn}$ and $\Phi_{mn}$ given by
\begin{eqnarray}
\phi_{mn} & = & \sum_{i} a^{i}_{m} (m_{mn}b^{i}_{n} - b^{i}_{m}
m_{mn}) \nonumber \\
\Phi_{mn} & = & \sum_{i} A^{i}_{m} (M_{mn}B^{i}_{n} - B^{i}_{m}
M_{mn})
\end{eqnarray}
The two form $d\rho$ will now be given by
\begin{equation}
d\rho = \sum_{i} d(a^{i}\otimes 1)d(b^{i}\otimes 1) + \sum_{i}
d(1\otimes A^{i})d(1\otimes B^{i}) \;\; ,
\end{equation}
with the image under $\pi$
\begin{equation}
\pi (d\rho ) = \sum_{i} [D, a^{i}\otimes 1 ][D, b^{i}\otimes 1] +
\sum_{i} [D, 1\otimes A^{i}][D, 1\otimes B^{i}] \;\; .
\end{equation}
Gauge invariance of the spinor action
 $< \Psi , (D + \pi(\rho )) \Psi >$ under the transformation $\Psi
\rightarrow$$ ^{g}\Psi = g\Psi$, where $g\in U({\cal A})\otimes U({\cal
B})$, demands that $\rho$ transforms inhomogenously such that
\begin{equation}
^{g}\rho = g\rho g^{\dagger} + gdg^{\dagger} \;\; ,
\label{gauget}
\end{equation}
where $g=g_{2}\otimes g_{4}$. This can be written as
\begin{eqnarray}
^g\rho = \{ \sum_{i} (g_{2}a^{i})d(b^{i}g_{2}^{\dagger})
-g_{2}(\sum_{i} a^{i}b^{i} -1)dg_{2}^{\dagger}\} \otimes 1
\;\;\;\;\;\;\;\;\;\;\;\;\;\;\;\;\;\;\;\;\;\;\;\;\;\nonumber\\ +
1\otimes \{ \sum_{i} (g_{4}A^{i})d(B^{i}g_{4}^{\dagger})
-g_{4}(\sum_{i} A^{i}B^{i} -1)dg_{4}^{\dagger}\}
\end{eqnarray}
so that gauge transformations can be defined directly on the
constituent elements by
\begin{eqnarray}
a^{i} {\rightarrow} ^{g}a^{i}=g_{2}a^{i} \;\;\;\;\;
A^{i} {\rightarrow} ^{g}A^{i}=g_{4}A^{i} \nonumber \\
b^{i} {\rightarrow} ^{g}b^{i}=b^{i}g_{2}^{\dagger} \;\;\;\;\;
B^{i} {\rightarrow} ^{g}B^{i}=B^{i}g_{4}^{\dagger}
\end{eqnarray}
if the constraints
\begin{equation}
\sum_{i} a^{i}b^{i} =1 \;\;\; {\rm and} \;\;\; \sum_{i} A^{i}B^{i} =1
\end{equation}
are imposed.

Gauge transformations can be expressed in the representation $\pi$
which from (\ref{gauget}) take the form
\begin{equation}
\pi ( ^{g}\rho ) = g\pi (\rho )g^{\dagger} + g[D, g^{\dagger}]
\end{equation}
and in component form become
\begin{eqnarray}
^{g}A_{2} = g_{2}A_{2}g_{2}^{\dagger} + g_{2}{\not \! \partial}
g_{2}^{\dagger} \;\;\;\;\;\;
^{g}A_{4} = g_{4}A_{4}g_{4}^{\dagger} + g_{4}{\not \! \partial}
g_{4}^{\dagger} \;\;\; m=1,2,3,4  \nonumber \\
^{g}(\phi_{mn}\otimes M_{mn}
 + m_{mn}\otimes \Phi_{mn} +m_{mn} \otimes
 M_{mn}) \;\;\;\;\;
\;\;\;\;\;\;\;\;\;\;\;\;\;\;\;\;\;\;\;\;\;\;\;\;\;
\;\;\;\;\;\;\;\;\;\;\;\;\;\;\;\;\;\;\;\;\;\;\;\;\;
 \nonumber \\ \;\;\;\;\;\; =  g_{2}\otimes g_{4}
(\phi_{mn}\otimes M_{mn} + m_{mn}\otimes \Phi_{mn} + m_{mn}\otimes
M_{mn})g_{2}^{\dagger}\otimes g_{4}^{\dagger} \;\;\; m{\not=}n
\label{gauge}
\end{eqnarray}
Thus $A_{2}$ and $A_{4}$ are indeed the $U(2)$ and $U(4)$ gauge fields
with the combination $(\phi_{mn}\otimes M_{mn} + m_{mn}\otimes
\Phi_{mn} + m_{mn}\otimes M_{mn})$ scalar fields transforming
covariantly, where $^{g}(m_{mn}\otimes M_{mn}) = m_{mn}\otimes M_{mn}$
in $D$.
The form of the scalar fields demonstrates that
$\phi_{mn}$ and $\Phi_{mn}$ represent independent fluctuations around
the vacuum state specified by $m_{mn}\otimes M_{mn}$.

The representation of the curvature $\pi (\Theta )$ requires a
determination of $\pi (d\rho )$ which, although a tedious calculation,
is a direct generalization of the computation presented in
{\cite{D}}. Thus, rather than outlining the detailed procedure we
will simply present the results. Expressed in terms of the gauge
fields, Higgs fields and auxiliary fields the diagonal elements of the
curvature can be written as
\begin{eqnarray}
\pi (\Theta )_{mm} = {1\over 2}\gamma^{\mu\nu} F^{m(2)}_{\mu\nu} +
{1\over 2}\gamma^{\mu\nu} F^{m(4)}_{\mu\nu}
+ \sum_{p\not= m}(|K_{mp}|^{2}(|\phi_{mp}\otimes M_{mp} +
m_{mp}\otimes \Phi_{mp} \nonumber \\ + m_{mp}\otimes M_{mp}|^{2}
+| m_{mp}\otimes M_{mp}|^{2})) -Y_{m}-X_{mm}
\end{eqnarray}
where
\begin{eqnarray}
X_{mm}  = \sum_{i} a^{i}_{m}{\not \! \partial}^{2} b^{i}_{m} +
\sum_{i} A^{i}_{m}{\not \! \partial}^{2} B^{i}_{m}
\hspace{7.5cm}
\nonumber \\
+(\partial^{\mu} A^{m}_{2\mu} + A^{m\mu}_{2}A^{m}_{2\mu}) +
(\partial^{\mu} A^{m}_{4\mu} + A^{m\mu}_{4}A^{m}_{4\mu})
- 2A^{m\mu}_{2}A^{m}_{4\mu} \;\; , \;\;\;\;\;\;\;\; \nonumber \\
Y_{m}  = \sum_{p\not= m} \sum_{i}
a^{i}_{m}|K_{mp}|^{2}|m_{mp}|^{2}\otimes |M_{mp}|^{2} b^{i}_{m} +
A^{i}_{m}|K_{mp}|^{2}|m_{mp}|^{2}\otimes |M_{mp}|^{2} B^{i}_{m}\;\; ,
\nonumber \\
F^{m(2)}_{\mu\nu}  =  \partial_{\mu} A^{m}_{2\nu} -\partial_{\nu}
A^{m}_{2\mu} + [A^{m}_{2\mu}, A^{m}_{2\nu}]
\;\;\; , \;\;\;\;
F^{m(4)}_{\mu\nu}  =  \partial_{\mu} A^{m}_{4\nu} -\partial_{\nu}
A^{m}_{4\mu} + [A^{m}_{4\mu}, A^{m}_{4\nu}]
\end{eqnarray}
and, for example, $|K_{mp}|^{2}=K_{mp}K_{pm}$.
The non-diagonal elements
of the curvature are given by $(m\not= n)$
\begin{eqnarray}
\pi (\Theta )_{mn} = -\gamma_{5} K_{mn}({\not \!
\partial}(\phi_{mn}\otimes M_{mn} + m_{mn}\otimes \Phi_{mn} )
\hspace{6.0cm}\nonumber \\
+ (A^{m}_{2}+A^{m}_{4})
(\phi_{mn}\otimes M_{mn} + m_{mn}\otimes \Phi_{mn} +
m_{mn}\otimes M_{mn})\hspace{3.0cm}\nonumber \\
-(\phi_{mn}\otimes M_{mn} + m_{mn}\otimes \Phi_{mn} +
m_{mn}\otimes M_{mn})(A^{n}_{2}+A^{n}_{4})) \hspace{3.0cm}\nonumber \\
+\sum_{p\not= m,n} K_{mp}K_{pn}((
\phi_{mp}\otimes M_{mp} + m_{mp}\otimes \Phi_{mp} +
m_{mp}\otimes M_{mp})\hspace{2.5cm}\nonumber \\
(\phi_{pn}\otimes M_{pn} + m_{pn}\otimes \Phi_{pn} +
m_{pn}\otimes M_{pn})
-m_{mp}m_{pn}\otimes M_{mp}M_{pn}) - X_{mn} \hspace{.2cm}
\end{eqnarray}
where
\begin{eqnarray}
X_{mn} = \sum_{i} \sum_{p\not= m,n}
K_{mp}K_{pn}\{a^{i}_{m}(m_{mp}m_{pn}\otimes M_{mp}M_{pn}b^{i}_{n}
 - b^{i}_{m}m_{mp}m_{pn}\otimes M_{mp}M_{pn}) \nonumber \\
+ A^{i}_{m}(m_{mp}m_{pn}\otimes M_{mp}M_{pn}B^{i}_{n}
-B^{i}_{m}m_{mp}m_{pn}\otimes M_{mp}M_{pn})\}.
\end{eqnarray}
Recall that we are
working in four dimensional Euclidean space so that the gamma matrices
employed satisfy: $\gamma_{\mu}^{\dagger} = -\gamma_{\mu},
\{\gamma_{\mu} , \gamma_{\nu}\} = -2\delta_{\mu\nu}$ and $ \gamma_{5}
= \gamma_{1}\gamma_{2}\gamma_{3}\gamma_{4}$. Note also that the
curvature is self-adjoint so that $\pi (\Theta )^{\dagger}_{mn} =
\pi (\Theta )_{nm}$. The Euclidean space action can now be determined
by exploiting (\ref{action}) and takes the form:
\begin{eqnarray}
I = -\int d^{4}x\sum_{m} Tr({1\over 4}F^{m(2)}_{\mu\nu}F^{m(2)\mu\nu}
+{1\over 4} F^{m(4)}_{\mu\nu}F^{m(4)\mu\nu} \hspace{7.0cm}\nonumber \\
-{1\over 2}|\sum_{p\not= m}(|K_{mp}|^{2}(|\phi_{mp}\otimes M_{mp} +
m_{mp}\otimes \Phi_{mp}
+ m_{mp}\otimes M_{mp}|^{2}
\hspace{3.5cm}\nonumber \\+| m_{mp}\otimes M_{mp}|^{2})) -Y_{m}-
X_{mm}|^{2} \hspace{3.5cm}\nonumber \\
+{1\over 2}\sum_{p\not= m,n} |K_{mp}|^{2}|({
\partial_{\mu}}(\phi_{mn}\otimes M_{mn} + m_{mn}\otimes \Phi_{mn} )
\hspace{5.5cm}\nonumber \\
+ (A^{m}_{2\mu}+A^{m}_{4\mu})
(\phi_{mn}\otimes M_{mn} + m_{mn}\otimes \Phi_{mn} +
m_{mn}\otimes M_{mn})\hspace{3.75cm}\nonumber \\
-(\phi_{mn}\otimes M_{mn} + m_{mn}\otimes \Phi_{mn} +
m_{mn}\otimes M_{mn})(A^{n}_{2\mu}+A^{n}_{4\mu}))|^{2}
\hspace{3.4cm}\nonumber \\
-{1\over 2}\sum_{n\not= m}\sum_{p\not= m,n} ||K_{mp}|^{2}((
\phi_{mp}\otimes M_{mp} + m_{mp}\otimes \Phi_{mp} +
m_{mp}\otimes M_{mp})\hspace{2.8cm}\nonumber \\
(\phi_{pn}\otimes M_{pn} + m_{pn}\otimes \Phi_{pn} +
m_{pn}\otimes M_{pn})
-m_{mp}m_{pn}\otimes M_{mp}M_{pn}) - X_{mn}|^{2})\hspace{1cm}
\label{baction}
\end{eqnarray}
where we normalize the trace such that $Tr 1=1$. Note that since
special unitary groups only will be considered, cross terms of the
field strengths have been ignored.

We can now address ourselves to the construction of an
$SU(2)_{L}\otimes SU(2)_{R}\otimes SU(4)$ partial unification model.
In order to achieve two independent $U(2)$ gauge symmetries, rather
than four, and to induce triplet Higgs fields, we will introduce the
permutation symmetries
\begin{eqnarray}
a^{i}_{1} = a^{i}_{2} \;\;\;\;\; a^{i}_{3} = a^{i}_{4} \nonumber \\
b^{i}_{1} = b^{i}_{2} \;\;\;\;\;\; b^{i}_{3} = b^{i}_{4}
\label{ident}
\end{eqnarray}
These are the identifications made by Chamseddine et al. {\cite
{D}} in their considerations on the left-right symmetric
electroweak model. However, rather than a graded tracelessness
condition on $\pi (\rho )_{1}$ we will simply impose the constraint
\begin{equation}
Tr(\pi (\rho )_{1}) = 0 \;\; ,
\end{equation}
reducing $U(2)_{L}\otimes U(2)_{R}$ to $SU(2)_{L}\otimes SU(2)_{R}$,
so avoiding the introduction of $U(1)$ factors. Since we want only one
$U(4)$ field all the copies in $\pi (\rho )_{2}$ must be identified. If
we were to choose identifications of the form (\ref{ident}) we would
be considering a model with $U(4)_{L}\otimes U(4)_{R}$ gauge symmetry.
The additional identifications which we must impose to avoid this are
analogous to the criteria required to yield an $SO(10)$ rather than an
$SU(16)$ symmetry in the $SO(10)$ models of Chamseddine and
Fr\"ohlich{\cite F}. That is, we are comparing the case of a model with
symmetry group $SU(2)_{L}\otimes SU(2)_{R}\otimes SU(4)_{L}\otimes
SU(4)_{R}$, which is a subgroup of $SU(16)$, with the Pati-Salam
partial unification which can be embedded as a maximal subgroup of
$SO(10)$.

To achieve a Higgs structure which will allow for the generation of a
large right handed neutrino mass at tree level we must consider, along
with direct identifications between space-times, the inclusion of a
conjugation symmetry. In this way we also introduce conjugate spinors
into the spinor representation $\Psi$. This is to be contrasted with
the electroweak case in which an additional conjugate set of fermions
was needed to produce the full range of allowed Yukawa couplings. As
$SO(1,3)$ and $SU(2)$ have conjugation matrices the conjugate spinors
can be written as
\begin{equation}
\psi^{c}_{L,R} = i\tau_{2} C {\overline \psi}^{T}_{L,R} \;\; ,
\end{equation}
where $C$ is the Dirac conjugation matrix and $i\tau_{2}$ the $SU(2)$
conjugation matrix. Since we require complex representations in the
$U(4)$ sector to be transformed to their complex conjugate, the charge
conjugation operator must be an outer automorphism on the
$U(4)$ algebra. The
conjugation symmetries that we impose in the $U(4)$ sector, then, are
given by
\begin{eqnarray}
A^{i}_{1} \leftrightarrow A^{i\ast}_{2} \;\;\;\;\;
A^{i\ast}_{3} \leftrightarrow A^{i}_{4}  \nonumber \\
B^{i}_{1} \leftrightarrow B^{i\ast}_{2} \;\;\;\;\;
B^{i\ast}_{3} \leftrightarrow B^{i}_{4}
\label{ident2}
\end{eqnarray}
where $A^{\ast}_{m}$ is the complex conjugate of $A_{m}$ corresponding
to the anti-representation, together with the identifications
\begin{eqnarray}
A_{1}^{i} = A_{4}^{i} \;\;\;\;\; A_{2}^{i} = A_{3}^{i}  \nonumber
\\
B_{1}^{i} = B_{4}^{i} \;\;\;\;\; B_{2}^{i} = B_{3}^{i}
\end{eqnarray}
so yielding a single $U(4)$ interaction.

Note that since all $SU(2)$ representations are real, the
identifications made in (\ref{ident}) can be equivalently considered
as conjugation symmetries. Thus, while all the space-time copies are
identified in the $U(4)$ sector the $SU(2)$ sector remains to
differentiate between the left and right regions of the model.
Futhermore within each region, left and right, a consistent
conjugation symmetry prevails between space-times. The added
complication in the $U(4)$ sector is, as before, related to its chiral
symmetry. The spinor $\Psi$ can thus be written as
\begin{equation}
\Psi =
\left (
\begin{array}{c}
\;\;\;\; \psi_{L} \;\;\;\; \\
\; i\tau_{2} C {\overline \psi}_{L}^{T} \; \\
\; i\tau_{2} C {\overline \psi}_{R}^{T} \; \\
\;\;\;\; \psi_{R} \;\;\;\;
\end{array}
\right )
\end{equation}
where the left and right handed assignments follow from imposing the
chirality condition
\begin{equation}
( \gamma_{5} \otimes \Gamma )\Psi = \Psi \;\; ,
\end{equation}
with $\Gamma = {\rm diag} (1,1,-1,-1)$,
introduced after the Wick rotation to Minkowski space.
Since all the elements in the
$U(4)$ sector are identified we will reduce $U(4) \rightarrow SU(4)$
by simply imposing $Tr(A_{4})=0$.

With our choice of symmetries between space-times the vector potential
$\pi (\rho )_{1}$ takes the form (suppressing the family matrices
$K_{mn}$ and the $\gamma_{5}$'s)
\begin{equation}
\pi (\rho )_{1} =
\left (
\begin{array}{cccc}
 \,\;\;\;\;\;\; A_{L} \;\;\;\; \;\;\;\;\;
\Delta_{L}^{(1)}\otimes M_{12} \;\;
\phi^{\prime}\otimes M_{13} \;\;\;\;\;\;
\phi\otimes M_{14}\;\;  \\
 \Delta_{L}^{(1)\dagger}\otimes M_{21} \;\;
\;\;\;\;\;\;\; {\overline A_{L}} \;\;\;\;\;\; \;\;\;
\phi^{\ast}\otimes M_{23} \;\;\;\;\;\;
\phi^{\prime\ast}\otimes M_{24}\;\;  \\
\;\; \phi^{\prime\dagger}\otimes M_{31} \;\;\;\;\;\;
\phi^{\ast\dagger}\otimes M_{32} \;\;
\;\;\;\;\;\;\; {\overline A_{R}} \;\;\;\;\;\;\; \;\;
\Delta_{R}^{(1)\dagger}\otimes M_{34}  \\
\;\; \phi^{\dagger}\otimes M_{41} \;\;\;\;\;\;
\phi^{\prime\ast\dagger}\otimes M_{42} \;\;\,
\Delta_{R}^{(1)}\otimes M_{43} \;\;
\;\;\;\;\;\; A_{R} \;\;\;\;\;\;\;\;\,
\end{array}
\right )
\end{equation}
where $A_{L}$ and $A_{R}$ are the $SU(2)_{L}$ and $SU(2)_{R}$ gauge
fields, $\Delta_{L}^{(1)}$ and $\Delta_{R}^{(1)}$ are singlets and
triplets in the respective groups and $\phi$ is a bi-doublet.
Similarly, $\pi (\rho )_{2}$ takes the form
\begin{equation}
\pi (\rho )_{2} =
\left (
\begin{array}{cccc}
 \;\;\;\;\;\;\; A_{4} \;\;\;\; \;\;\;
\, m_{12}\otimes\Delta^{(2)} \;\;
m_{13}\otimes\Delta^{(2)} \;\;
\; m_{14}\otimes\Sigma\;\;\;\;  \\
 m_{21}\otimes\Delta^{(2)\dagger} \;\;
\;\;\;\;\; {\overline A_{4}} \;\;\;\; \;\;\;
\;\, m_{23}\otimes \Sigma \;\;\;\;\;
m_{24}\otimes\Delta^{(2)\dagger}  \\
 m_{31}\otimes\Delta^{(2)\dagger}   \;\;
\; m_{32}\otimes\Sigma \;\;\;\;\;
\;\;\;\;\;\, {\overline A_{4}} \;\;\;\; \;\;\;
m_{34}\otimes\Delta^{(2)\dagger}  \\
 \; m_{14}\otimes\Sigma \;\;\;\;\;
m_{42}\otimes\Delta^{(2)} \;\;
m_{43}\otimes\Delta^{(2)} \;\;
\;\;\;\;\; A_{4} \;\;\;\;\;\;\;
\end{array}
\right )
\end{equation}
{}From the space-time symmetries the constituent fields transform as
\begin{eqnarray}
\Delta^{(2)} \sim {\bf 4\times 4} & = & {\bf 6+10} \nonumber \\
\Sigma \sim {\bf 4\times {\overline 4}} & = & {\bf 1+15}
\end{eqnarray}
under $SU(4)$. We know the covariant form taken by Higgs scalars under
a general gauge transformation from (\ref{gauge}). Consequently, the
Higgs fields which enter the model will transform under
$SU(2)_{L}\otimes SU(2)_{R}\otimes SU(4)$ as:
\begin{eqnarray}
\Delta_{L} = \Delta_{L}^{(1)}\otimes M_{12} + m_{12}\otimes\Delta^{(2)}
& \sim & {\bf (3,1,6) + (3,1,10) + (1,1,6) + (1,1,10)}\nonumber \\
\Delta_{R} = \Delta_{R}^{(1)}\otimes M_{43} + m_{43}\otimes\Delta^{(2)}
& \sim & {\bf (1,3,6) + (1,3,10) + (1,1,6) + (1,1,10)}\nonumber \\
\Phi = \phi\otimes M_{14} + m_{14}\otimes \Sigma \;\;\;\;
& \sim & {\bf (2,2,1)
+ (2,2,15)} \nonumber \\
\Phi^{\prime} = \phi^{\prime}
\otimes M_{13} + m_{13}\otimes\Delta^{(2)} & \sim &
{\bf (2,2,6) + (2,2,10)}.
\label{reps}
\end{eqnarray}
The other entries follow from the inter-space-time symmetries. This in
turn breaks the Higgs field degeneracy in the vector potential.
Embedding our partial unification model into $SO(10)$ we see
immediately that our Higgs fields transform as components of relevant
Higgs representations often chosen for such models, e.g. $\bf 10, 120,
126$ and $\bf 210$. We could easily obtain other components by
different choices of symmetries between space-times. Nevertheless,
with the choice of symmetries taken we have generated the Higgs
components which are required to have non-zero vacuum expectation
values for a non-minimal model.

With the product structure between the $SU(2)$ and $SU(4)$ sectors
made manifest in the form of the Higgs scalars we must correspondingly
realize this mathematical structure in the spinors, which take the form
\begin{equation}
\psi_{L,R} =
\left [
\begin{array}{l}
\; u \; \\
\; d \;
\end{array}
\right ]
\otimes (a,b,c,d)
\end{equation}
where the column $[u,d]$ indicates valency and the spin-zero row
$(a,b,c,d)$ indicates colour degrees of freedom{\cite G}.
This is to be compared
with the physical realization given by (\ref{mulstr}). From the spinor
action $<\Psi , (D + \pi (\rho ))\Psi >$
we see that we can generate the
Yukawa couplings:
\begin{eqnarray}
{\cal L}_{Y} =
K_{14}{\overline \Psi}_{L} (\Phi + <\Phi >) \Psi_{R} +
K_{32}{\overline \Psi}_{L} \tau_{2}(\Phi^{\ast} + <\Phi^{\ast} >)
^{\dagger}
\tau_{2} \Psi_{R} \hspace{1.5cm}\nonumber \\ -
iK_{21}\Psi_{L}^{T}\tau_{2} C^{-1} (\Delta_{L} + <\Delta_{L}
>)^{\dagger} \Psi_{L}
-iK_{34}\Psi_{R}^{T}\tau_{2} C^{-1} (\Delta_{R} + <\Delta_{R}
>)^{\dagger} \Psi_{R} +
{\rm H.c.}
\end{eqnarray}
where the Hermitian conjugates emerge automatically from
the self-adjointness of $\pi (\rho )$. Note that the implied non-gauge
invariant couplings have disappeared from the model by setting the
vacuum expectation values of such Higgs' to zero. The coupling to
$\Phi$ will again be responsible for the usual quark and lepton
masses. However, we now have a tree level coupling which can yield
large right handed neutrino masses by the see-saw mechanism, i.e. the
coupling producing Majorana masses given by the components $\bf
(3,1,10)$ and $\bf (1,3,10)$ of $\Delta_{L}$ and $\Delta_{R}$
respectively{\cite G}. Coupling to the conjugate bi-doublet Higgs is a
natural consequence of the introduction of conjugation symmetries.

The true viability of the model is dependent on the survival of the
Higgs potential once suitable vacuum expectation values for the Higgs
scalars have been chosen. This corresponds to the elimination of
unwanted components from (\ref{reps}), where we will take the only
components with non-zero vacuum expectation values to be
$\Delta_{L}\sim {\bf (3,1,10)},\; \Delta_{R}\sim {\bf (1,3,10)}$ and
$\Phi \sim {\bf (2,2,1)}$. The symmetry breaking scheme will then take
the form:
\newline
\newcounter{cms}
\setlength{\unitlength}{1mm}
\begin{picture}(170,25)(0,-1)
\put(25,7){\makebox(0,0){$SU(2)_{L}\otimes SU(2)_{R}\otimes SU(4)$}}
\put(56.5,15){\makebox(0,0){$<\Delta_{R} >=v_{R}$}}
\put(90,7){\makebox(0,0){$SU(2)_{L}\otimes SU(3)_{C}\otimes
U(1)_{Y}$}}
\put(121.5,15){\makebox(0,0){$<\Phi >\not= 0$}}
\put(146.5,7){\makebox(0,0){$SU(3)_{C}\otimes U(1)_{Q}$}}
\thicklines
\put(50,7){\vector(1,0){15}}
\put(115.5,7){\vector(1,0){15}}
\end{picture}
where we have the expectation value hierachy $<\Delta_{L}>\,
 \ll \,<\Phi>\, \ll \,
<\Delta_{R}>$. The form of the vacuum
expectation values is dictated by the requirement that $U(1)_{Q}$
survive so that only charge zero components can be non-zero. For a
fractionally charged quark model in which the gluons also remain
chargeless
the charge operator $Q$ is given by{\cite G}
\begin{equation}
Q = I_{3L} + I_{3R} + {1/2}
\left (
\begin{array}{rrrr}
\; {1/3} \;\; 0 \;\; 0 \;\; 0 \; \\
\; 0 \;\;  {1/3} \;\; 0 \;\; 0 \; \\
\; 0 \;\; 0 \;\; {1/3} \;\; 0 \; \\
 0 \;\; 0 \;\;\;\; 0 \, {-1} \;
\end{array}
\right )
\end{equation}
so that the Higgs vacuum expectation values become
\begin{eqnarray}
<\Delta_{R} > = v_{R}
\left (
\begin{array}{cc}
\; 0 \;\; 1 \; \\
\; 0 \;\; 0 \;
\end{array}
\right )
\otimes
\left (
\begin{array}{rrrr}
\; 0 \;\; 0 \;\; 0 \;\; 0 \; \\
\; 0 \;\; 0 \;\; 0 \;\; 0 \; \\
\; 0 \;\; 0 \;\; 0 \;\; 0 \; \\
\; 0 \;\; 0 \;\; 0 \;\; 1 \; \\
\end{array}
\right )
= v_{R} S_{2}\otimes S_{4} \hspace{1.1cm}
\nonumber \\
<\Delta_{L} > = v_{L}
\left (
\begin{array}{cc}
\; 0 \;\; 1 \; \\
\; 0 \;\; 0 \;
\end{array}
\right )
\otimes
\left (
\begin{array}{rrrr}
\; 0 \;\; 0 \;\; 0 \;\; 0 \; \\
\; 0 \;\; 0 \;\; 0 \;\; 0 \; \\
\; 0 \;\; 0 \;\; 0 \;\; 0 \; \\
\; 0 \;\; 0 \;\; 0 \;\; 1 \;\\
\end{array}
\right )
= v_{L} S_{2}\otimes S_{4} \hspace{1.1cm}
\nonumber \\
<\Phi > =
\left (
\begin{array}{cc}
\; u_{1} \;\; 0 \; \\
\; 0 \;\; u_{2} \;
\end{array}
\right )
\otimes
\left (
\begin{array}{rrrr}
\; 1 \;\; 0 \;\; 0 \;\; 0 \; \\
\; 0 \;\; 1 \;\; 0 \;\; 0 \; \\
\; 0 \;\; 0 \;\; 1 \;\; 0 \; \\
\; 0 \;\; 0 \;\; 0 \;\; 1 \;
\end{array}
\right )
=
\left (
\begin{array}{cc}
\; u_{1} \;\; 0 \; \\
\; 0 \;\; u_{2} \;
\end{array}
\right )
\otimes I_{4}
\end{eqnarray}
We can now determine the independent contributions from the auxiliary
fields which must be eliminated. The $X$ and $Y$ fields are given by
\begin{eqnarray}
Y_{1} = |K_{12}|^{2} (\sum_{i} a^{i}_{1}
|v_{L}|^{2}S_{2}S_{2}^{\dagger}\otimes
S_{4}b^{i}_{1} + A^{i}|v_{L}|^{2}
S_{2}S_{2}^{\dagger}\otimes S_{4}B^{i} ) \hspace{5cm}\nonumber \\
+ |K_{14}|^{2} (\sum_{i} a^{i}_{1}
\left (
\begin{array}{ll}
 |u_{1}|^{2} \;\;\;\; 0 \;\; \\
\;\; 0 \;\;\;\; |u_{2}|^{2}
\end{array}
\right )
\otimes I_{4} b^{i}_{1} + A^{i}
\left (
\begin{array}{ll}
 |u_{1}|^{2} \;\;\;\; 0 \;\; \\
\;\; 0 \;\;\;\; |u_{2}|^{2}
\end{array}
\right )
\otimes I_{4} B^{i} ) \hspace{2.8cm}\nonumber \\
\nonumber \\
Y_{3}=
|K_{34}|^{2}(\sum_{i} a^{i}_{3} |v_{R}|^{2}S_{2}^{\dagger}S_{2}\otimes
S_{4} b^{i}_{3} + A^{i\ast} |v_{R}|^{2}S_{2}^{\dagger}S_{2}\otimes
S_{4} B^{i\ast} ) \hspace{4.7cm}\nonumber \\
 +|K_{32}|^{2} ( \sum_{i} a^{i}_{3}
\left (
\begin{array}{ll}
 |u_{1}|^{2} \;\;\;\; 0 \;\; \\
\;\; 0 \;\;\;\; |u_{2}|^{2}
\end{array}
\right )
\otimes I_{4} b^{i}_{3} + A^{i\ast}
\left (
\begin{array}{ll}
 |u_{1}|^{2} \;\;\;\; 0 \;\; \\
\;\; 0 \;\;\;\; |u_{2}|^{2}
\end{array}
\right )
\otimes I_{4} B^{i\ast}) \hspace{2.5cm}\nonumber \\ \nonumber \\
X_{12}=X_{34}=X_{23}=X_{14}=0 \hspace{10.3cm}\nonumber \\ \nonumber \\
X_{13} =\sum_{i} K_{12}K_{23}\{ a_{1}^{i} v_{L}u_{2}(S_{2}\otimes
S_{4}b_{3}^{i} - b_{1}^{i}S_{2}\otimes S_{4})
+ A^{i}v_{L}u_{2}(S_{2}\otimes S_{4}B^{i\ast} - B^{i}S_{2}\otimes
S_{4})\} \nonumber \\
+ K_{14}K_{43}\{ a_{1}^{i} v_{R}u_{1}(S_{2}\otimes
S_{4}b_{3}^{i} - b_{1}^{i}S_{2}\otimes S_{4})
+ A^{i}v_{R}u_{1}(S_{2}\otimes S_{4}B^{i\ast} -
B^{i}S_{2}\otimes
S_{4})\}\hspace{.2cm}
\end{eqnarray}
where the others follow from the permutation symmetries. Clearly, the
field $X_{13}$ is auxiliary and thus must be eliminated . The
independent contributions from the $Y$ fields can be easily found
where, for example, $Y_{1}$ can be rewritten as
\begin{eqnarray}
Y_{1} = |K_{12}|^{2}(2|v_{L}|^{2}S_{2}S_{2}^{\dagger}\otimes S_{4}
-(\sum_{i} a_{1}^{i}|v_{L}|^{2}(S_{2}^{\dagger}S_{2}\otimes
S_{4}b_{1}^{i} -b^{i}_{1}S_{2}^{\dagger}S_{2}\otimes S_{4}))
\hspace{4cm}\nonumber\\
- (\sum_{i} A^{i}|v_{L}|^{2}(S_{2}S_{2}^{\dagger}\otimes
\left (
\begin{array}{rrrr}
\; 1 \;\; 0 \;\; 0 \;\; 0 \; \\
\; 0 \;\; 1 \;\; 0 \;\; 0 \; \\
\; 0 \;\; 0 \;\; 1 \;\; 0 \; \\
\; 0 \;\; 0 \;\; 0 \;\; 0 \;
\end{array}
\right )
B^{i} -B^{i}S_{2}S_{2}^{\dagger}\otimes
\left (
\begin{array}{rrrr}
\; 1 \;\; 0 \;\; 0 \;\; 0 \; \\
\; 0 \;\; 1 \;\; 0 \;\; 0 \; \\
\; 0 \;\; 0 \;\; 1 \;\; 0 \; \\
\; 0 \;\; 0 \;\; 0 \;\; 0 \;
\end{array}
\right )
)))
\hspace{3.5cm}\nonumber \\
+|K_{14}|^{2}(2
\left (
\begin{array}{ll}
 |u_{1}|^{2} \;\;\;\; 0 \;\; \\
 \;\; 0 \;\;\;\; |u_{2}|^{2}
\end{array}
\right )
\otimes I_{4} -(\sum_{i} a_{1}^{i} (
\left (
\begin{array}{cc}
 |u_{2}|^{2} \;\;\;\; 0  \;\; \\
\;\; 0 \;\;\;\; |u_{1}|^{2}
\end{array}
\right )
\otimes I_{4}b_{1}^{i} - b_{1}^{i}
\left (
\begin{array}{rr}
 |u_{2}|^{2} \;\;\;\;\; 0 \;\; \\
\;\; 0 \;\;\;\;\; |u_{1}|^{2}
\end{array}
\right )
\otimes I_{4})))\hspace{.7cm} \nonumber
\end{eqnarray}
with the others following similarly. We therefore have the
non-independent contributions
\begin{eqnarray}
2(|K_{12}|^{2}|v_{L}|^{2}S_{2}S_{2}^{\dagger}\otimes S_{4}
+ |K_{14}|^{2}
\left (
\begin{array}{rr}
 |u_{1}|^{2} \;\;\;\; 0 \;\; \\
\;\; 0 \;\;\;\; |u_{2}|^{2}
\end{array}
\right )
\otimes I_{4}) \nonumber \\\nonumber \\
=2(|K_{12}|^{2}|m_{12}\otimes M_{12}|^{2} + |K_{14}|^{2}|m_{14}\otimes
M_{14}|^{2})\hspace{.7cm}
\end{eqnarray}
where the remainder is eliminated. Note that the $X_{mm}$ are also
eliminated. The Higgs potential can thus be shown to take the form
\begin{eqnarray}
V=
(Tr|K_{12}|^{4}-(Tr|K_{12}|^{2})^{2})||\Delta_{L}+m_{12}\otimes
M_{12}|^{2} - |m_{12}\otimes M_{12}|^{2}|^{2} \hspace{.4cm}\nonumber \\
+(Tr|K_{14}|^{4}-(Tr|K_{14}|^{2})^{2})||\Phi + m_{14}\otimes
M_{14}|^{2} - |m_{14}\otimes M_{14}|^{2}|^{2} \hspace{.35cm}
\nonumber \\
+(Tr|K_{23}|^{4}-(Tr|K_{23}|^{2})^{2})||\Phi^{\ast} + m_{23}\otimes
M_{23}|^{2} - |m_{23}\otimes M_{23}|^{2}|^{2} \hspace{.2cm}\nonumber \\
+(Tr|K_{43}|^{4}-(Tr|K_{43}|^{2})^{2})||\Delta_{R}+m_{43}\otimes
M_{43}|^{2} - |m_{43}\otimes M_{43}|^{2}|^{2} \hspace{.1cm}
\end{eqnarray}
and thus survives.

Looking more closely at the fermionic action it follows from the
quark-lepton unification that the same family mixing matrix will
operate on the $u$ and $d$ quarks. This does not occur in the $SU(5)$
and standard model examples previously considered since an additional
set of spinors must be introduced so that the $u$ quark may attain a
non-zero mass{\cite D}.
This also allows for different mixing matrices and thus
the existence of a Cabibbo angle. Quark-lepton unification eliminates
the need to introduce an additional set of spinors and also,
therefore, different mixing matrices. This was the dilema faced in the
$SO(10)$ model of Chamseddine and Fr\"ohlich{\cite F}
for which introducing
singlet spinors resolved the problem. Compelled by the success of the
model we have constructed so far we will take a different approach to
this problem. For models without a right handed neutrino no mixing
occurs in the neutrino sector because the mass matrix is identically
zero. Quark-lepton unification implies breaking this neutrino
degeneracy. Note also that we
now have an additional degree of freedom from introducing a bi-module.
With the addition of $SU(4)$ we will thus correspondingly consider the
mixing matrices to take the form
\begin{equation}
K_{mn}={\rm diag}(f_{\gamma\delta},f_{\alpha\beta})_{mn} \;\; ,
\end{equation}
acting on $d^{\alpha}$ and $u^{\gamma}$ in the multiplet structure of
(\ref{mulstr}). This is analogous to the
additional mixing allowed in other models by giving the $u$ quarks
non-zero masses as well as providing mixing among neutrinos. In this
way $u$ quarks and neutrinos attain additional structure on the same
footing, which is consistent with the concept of neutrinos being the
fourth up quark.

It is straightforward now to write down the full fermionic and bosonic
action from (\ref{faction}) and (\ref{baction}),
where consistent normalizations for the kinetic energies can be
accomodated by an appropriate rescaling. Since we have a bi-module
structure separating the introduction of $SU(2)$ and $SU(4)$ we are
free to implement different coupling strengths in these sectors.
We thus have an acceptable model for which many
phenomenonlogical parameters may be adjusted to yield results close to
the experimental values. Importantly, we have avoided extending our
Higgs sector beyond that needed to produce the required symmetry
breaking pattern.

\section{Conclusion}

By generalizing previous approaches to include a non-trivial extension
by way of a bi-module we have been able to formulate a model which can
yield tree level masses to neutrinos and so avoid the quantization
problem and the inclusion of exotic
fermions. Furthermore, we have broken
the Higgs field degeneracy inherent in the vector potential thus
enlarging our choice of scalar fields. The symmetry breaking scales,
corresponding to the ``distance" between copies of space-time, can now
find a geometrical basis. This is particularly important for models
such as this which incorporate intermediate scales.
In the absence of a quantization mechanism
or embedding of supersymmetry into non-commutative geometry our
approach yields a consistent formulation as well as utilizing more
fully the freedom provided by this mathematical framework.

\section{Acknowledgements}

It is a pleasure to thank J. Choi and V. Sivananthan for
helpful discussions. BEH would like to acknowledge the support of an
Australian Postgraduate Research Award.

\end{document}